\documentclass[twocolumn,preprintnumbers]{revtex4}
\usepackage{graphicx}
\usepackage{amsmath}
\usepackage{amssymb}
\usepackage{amsthm}
\usepackage{amsfonts}
\usepackage{enumerate}
\usepackage{centernot}
\usepackage{tikz}

\input{tcilatex}
\begin{document}

\title{Emergence of Topological Nodal Lines and Type II Weyl Nodes in Strong
Spin--Orbit Coupling System InNbX$_{2}$(X=S,Se)}
\author{Yongping Du$^{1}$, Xiangyan Bo$^{2}$, Di Wang$^{2}$, Er-jun Kan$^{1}$%
, Chun-Gang Duan$^{3,4}$, Sergey Y. Savrasov$^{5\ast }$, Xiangang Wan$%
^{2,6\ast }$}
\affiliation{$^{1}$Department of Applied Physics, Nanjing University of Science and
Technology, Nanjing, Jiangsu 210094, China\\
$^{2}$National Laboratory of Solid State Microstructures and Department of
Physics, Nanjing University, Nanjing 210093, China\\
$^{3}$Key Laboratory of Polar Materials and Devices, Ministry of Education,
East China Normal University, Shanghai, 200241, China\\
$^{4}$Collaborative Innovation Center of Extreme Optics, Shanxi University,
Taiyuan, Shanxi 030006, China\\
$^{5}$Department of Physics, University of California, Davis, One Shields
Avenue, Davis, California 95616, USA\\
$^{6}$Collaborative Innovation Center of Advanced Microstructures, Nanjing
University, Nanjing 210093, China}
\email{savrasov@physics.ucdavis.edu}
\email{xgwan@nju.edu.cn}

\begin{abstract}
Using first--principles density functional calculations, we systematically
investigate electronic structures and topological properties of InNbX$_{2}$
(X=S, Se). In the absence of spin--orbit coupling (SOC), both compounds show
nodal lines protected by mirror symmetry. Including SOC, the Dirac rings in
InNbS$_{2}$ split into two Weyl rings. This unique property is distinguished
from other dicovered nodal line materials which normally requires the
absence of SOC. On the other hand, SOC breaks the nodal lines in InNbSe$_{2}$
and the compound becomes a type II Weyl semimetal with 12 Weyl points in the
Brillouin Zone. Using a supercell slab calculation we study the dispersion
of Fermi arcs surface states in InNbSe$_{2}$, we also utilize\textbf{\ }a
coherent potential approximation to\ probe their tolernace to the surface
disorder effects.\ The quasi two--dimensionality and the absence of toxic
elements makes these two compounds an ideal experimental platform for
investigating novel properties of topological semimetals.
\end{abstract}

\date{\today }
\maketitle

\section{I. Introduction}

In the past few years, topological semimetals, such as Weyl semimetals (WSM) 
\cite{Weyl-1,Weyl-2}, Dirac semimetals (DSM) \cite%
{DSM-1,nagaosa,zhang,Na3Bi,Cd3As2,Cava-BaAgBi,BaAgBi} and Nodal Line
semimetals (NLS)\cite{NSL-1,NSL-2,NSL-3,NSL-4}, have received tremendous
research interest. In a Weyl semimetal, the electrons around Weyl points,
which are the crossing points of two non--degenerate linearly dispersing
energy bands, behave exactly like Weyl fermions\cite{Weyl-1,Weyl-2}. With
definite chirality, each Weyl point can be considered as a topologically
protected charge, thus extending classification of topological phases of
matter beyond insulators\cite{Weyl-1,Weyl-2}. Weyl points are extremely
robust against weak perturbations and can only be annihilated when pairs of
Weyl points with opposite topological charge meet with each other. Then the
system opens a gap evolving into either a normal insulator or an Axion
insulator\cite{Weyl-1,axion}.

One of the most remarkable properties of WSMs is the existence of
topological surface states in a form of Fermi arcs\cite{Weyl-1}. This serves
as an unambiguous evidence to identify this state of matter. A great number
of other exotic phenomena has also been proposed for WSMs: a highly
anisotropic negative magnetoresistance related to chiral anomaly effect\cite%
{Chiral-1,Chiral-2}, a topological response\cite{Weyl-response}, unusual
non--local transport properties\cite{non-local}, novel quantum oscillations
from Fermi arcs\cite{Quantum-oscillations}, etc.

A further classification here has been given to distinguish WSMs whose bulk
Fermi surfaces shrink to Weyl points (called type I), or to exotic
hyperboloid surfaces (called type II), where the cones are titled and induce
a finite density of states at the nodal point\cite{WTe2}. Due to the tilted
nature of the nodes, the low energy excitations break Lorentz invariance,
cause absence of the chiral anomaly at certain magnetic--field angles,
magnetic breakdown and novel Klein tunneling \cite{WTe2,Type-II-2}.

\begin{figure*}[tbh]
\center\includegraphics[width=7in, height=2.3in]{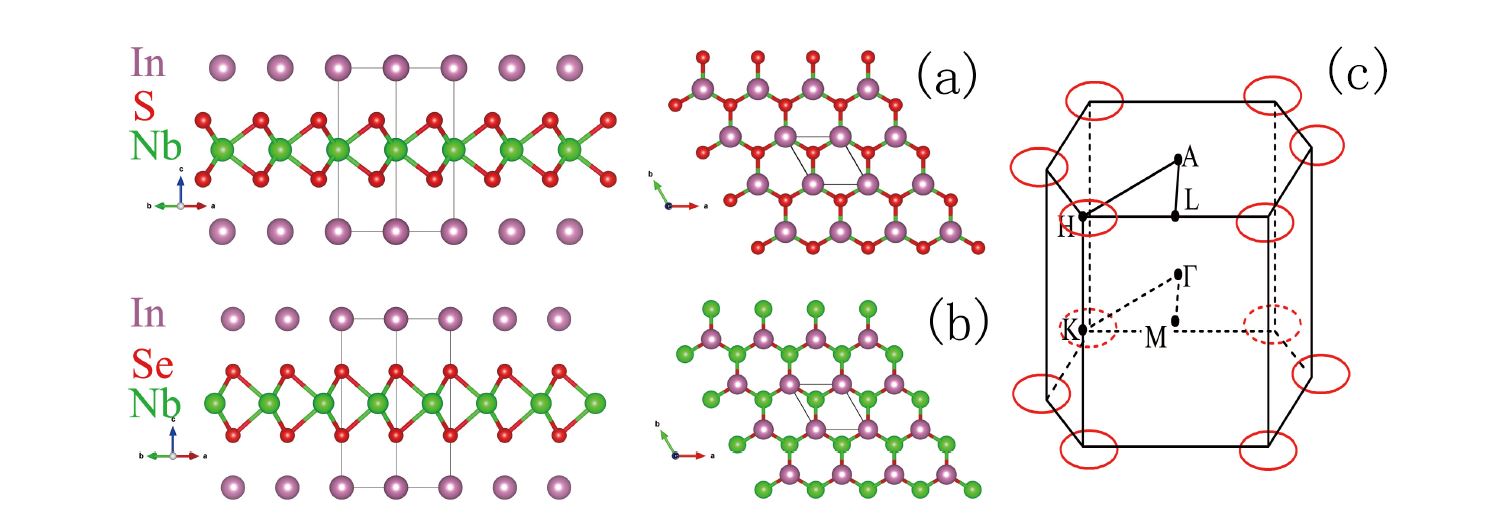}
\caption{(color online). (a) Side view (left) and top view (right) of
crystal structure of InNbS$_{2}$; (b) Side view (left) and top view (right)
of crystal structure of InNbSe$_{2}$. Green, violet, and red spheres
represent Nb, In, and S (Se) atoms, respectively. (c)The schematic of the
nodal lines in InNbX$_{2}$ (X=S,Se).}
\end{figure*}

There has been a great progress in searching for signatures of WSMs in real
materials. Starting from the original proposal on pyrochlore iridates \cite%
{Weyl-1}, several systems, such, e.g., as HgCr$_{2}$Se$_{4}$ \cite{HgCr2Se4}%
, TaAs\cite{TaAs,TaAs-hasan}, WTe$_{2}$\cite{,WTe2}, NbP\cite{S2}, TaP\cite%
{S3}, NbAs\cite{S4}, MoTe$_{2}$\cite{8-exp-MoTe2}, MoP$_{2}$, WP$_{2}$\cite%
{S6}, LaAlGe\cite{S7}, etc. have been predicted to exhibit WSM behavior. A
large amount of recent experimental work has been devoted to study
properties of\textbf{\ }TaAs family \cite%
{S2,S3,S4,TaAs-exp-1,TaAs-exp-2,TaAs-exp-3,TaAs-exp-5}. In many of the
proposed materials, however, the Weyl points do not exactly cross but only
close to the Fermi level, and, in addition, there also are trivial Fermi
states. The contribution from these trivial states significantly complicates
the analysis of topological surface states and their novel transport
behavior.

In addition to WSMs, a three--dimensional (3D) DSM \cite%
{DSM-1,nagaosa,zhang,Na3Bi,Cd3As2,Cava-BaAgBi,BaAgBi} has also been
proposed. The Dirac points in DSM are four--fold degenerate, and can be
viewed as a merge of two Weyl fermions with opposite chirality in the
Brillouin zone (BZ). The Dirac points usually require a protection by time
reversal, inversion and additional crystal symmetry \cite%
{nagaosa,zhang,Na3Bi,Cd3As2}.

Different from WSMs and DSMs which have finite numbers of band touching
points in the BZ, a third topological semimetal, NLS has a whole crossing
line in momentum space\cite{NSL-1,NSL-2,NSL-3,NSL-4}. Same as 3D DSM, NLS
also needs crystal symmetry to stabilize its band crossing line \cite%
{zhang,fuliang}. The most exotic property of NLS is its two dimensional (2D)
drumhead--like\ surface state\cite%
{NSL-1,NSL-2,NSL-3,NSL-4,CaP3,BaSn2,CaTe,NSL-mat-3,NSL-mat-4,NSL-mat-5,TlTaSe2,PbTaSe2,Ca3P2}%
. It has been speculated that this special state may realize
high--temperature superconductivity \cite{NSL-app-1,NSL-app-2}.

Several materials have been predicted to be topological NLSs \cite%
{NSL-1,NSL-2,NSL-3,NSL-4,CaP3,BaSn2,CaTe,NSL-mat-3,NSL-mat-4,NSL-mat-5,TlTaSe2,PbTaSe2,Ca3P2}%
. However, most of these predictions are based on calculations without
spin--orbit coupling (SOC) \cite%
{NSL-2,NSL-3,NSL-4,CaP3,BaSn2,CaTe,NSL-mat-3,NSL-mat-4,NSL-mat-5,Ca3P2},
inclusion of which leads normally to gapping out the nodal line \cite%
{NSL-2,NSL-3,NSL-4,CaP3,BaSn2,CaTe,NSL-mat-3,NSL-mat-4,NSL-mat-5,Ca3P2}.
Only in a few systems, this was found to be not the case, where TlTaSe$_{2}$
and PbTaSe$_{2}$ are predicted to remain NLS behavior in calculations with
SOC \cite{TlTaSe2,PbTaSe2}.

Formally, topological semimetals exist only for 3D systems. However,
quasi--2D layered materials that are easier to cleave and study their
surface electronic structures are more favorable from the experimental
perspective. Also, toxic elements like As, P, Tl and Hg found in many of the
discovered materials create additional complications. Thus, searching for
new topological systems and finding ways to remove the effects of trivial
states meanwhile preserving contributions from topological Fermi arcs, are
important problems of this emergent field of condensed matter physics.

In this work, we use first--principles calculations based on density
functional theory (DFT) in its generalized gradient approximation (GGA) \cite%
{GGA} to predict that InNbS$_{2}$ and InNbSe$_{2}$ show nodal lines and Weyl
semimetal behavior, respectively. Without considering SOC, both of them are
NLSs, and the band crossing lines are formed by four--fold degenerate Dirac
points. The nodal lines, which are located around H point in the $k_{z}=\pi $
plane (i.e. L-H-A plane) in the BZ, are protected by the mirror symmetry.
Including SOC, the four--fold degenerate nodal line in InNbS$_{2}$ splits
into two Weyl type nodal lines which are again protected by the mirror
symmetry. On the other hand, the SOC changes InNbSe$_{2}$ to a type II WSM
for which we predict the Fermi arc surface states to appear on a easily
cleavable (001) Indium terminated surface. Using a combination of DFT with
Coherent Potential Approximation (CPA)\ we also simulate the effects of
surface disorder to study the robustness of the Fermi arcs in this system.
Our theoretical work shows that the InNbS$_{2}$ and InNbSe$_{2}$ are very
promising materials for studying NLSs and WSMs, respectively.

\section{II. Crystal Structure}

The crystal structures of InNbX$_{2}$ (X=S, Se) belong to space group $P%
\overline{6}m2$ (NO. 187) which is non--centrosymmetric\cite%
{cry-1,cry-2,cry-3}. The In layer is intercalated between two niobium
dichalcogenides layers. As shown in Fig. 1(a) and Fig. 1(b), the mirror
plane is located at In layer or Nb layer. As discussed later in our work,
this mirror plane plays a key role in protecting the nodal line. As shown in
Fig.1, the In atoms in InNbS$_{2}$ are aligned with Nb atoms in the vertical
direction, while in InNbSe$_{2}$, they are aligned with Se atoms. This
difference in the lattice structure results in different topological
features for these two compounds. 
\begin{figure*}[tbh]
\center\includegraphics[width=7in, height=6in]{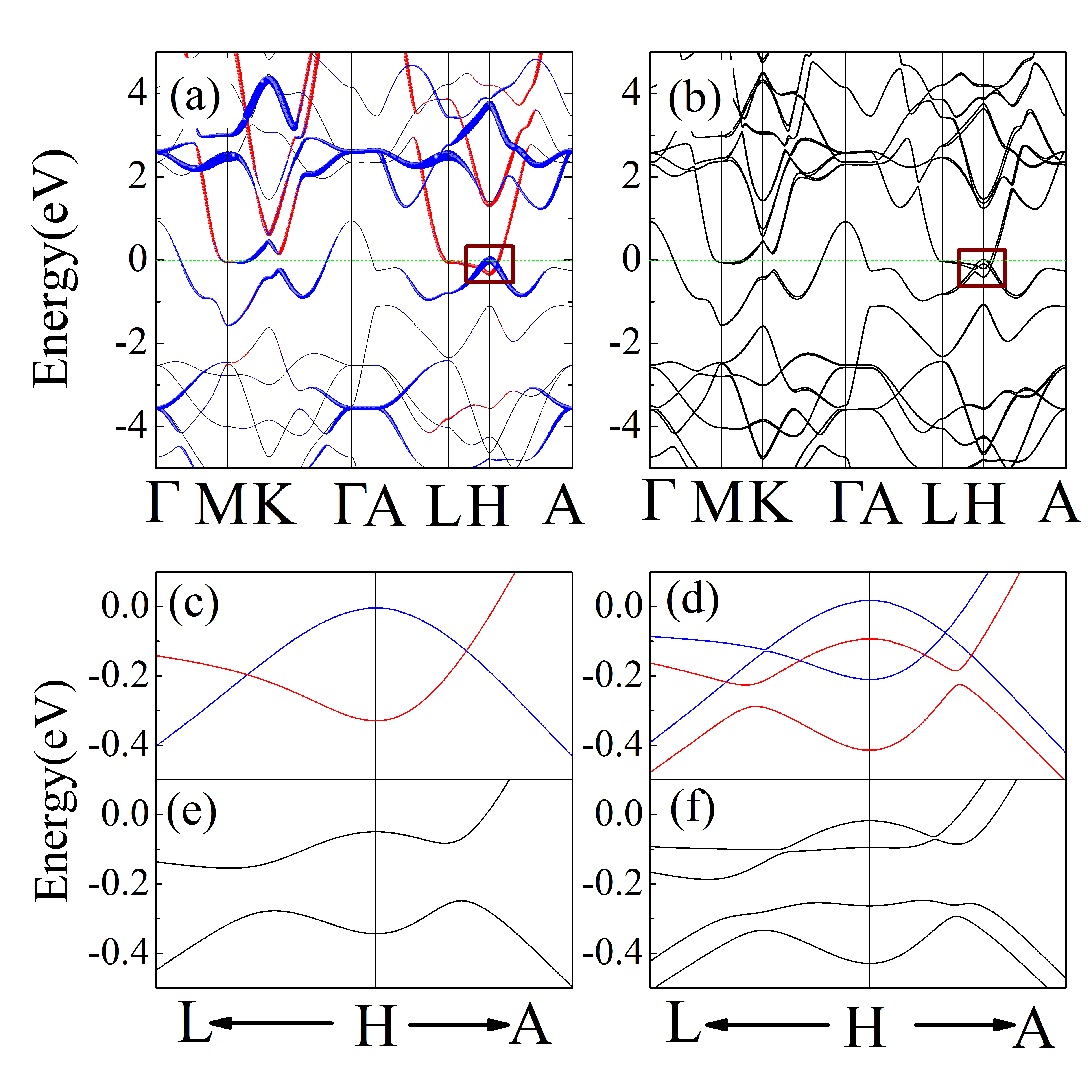}
\caption{(color online). (a) Calculated bulk band strucutre of InNbS$_{2}$
without SOC. The weights of In-6$p_{x}$/6$p_{y}$ (Nb-5$d_{x^{2}-y^{2}}$/5$%
d_{xy}$) states are proportional to the width of red (blue) curves. (b) Bulk
band structure of InNbS$_{2}$ calculated with the inclusion of SOC. (c)
Closeup of band structure around H point as marked in panel (a). The red
line denote the In-6$p_{x}$/6$p_{y}$ states which belong to irreducible
representation $\Gamma _{2}$ of space group, while the blue one is the state
of Nb-5$d_{x^{2}-y^{2}}$/5$d_{xy}$ belonging to irreducible representation $%
\Gamma _{1}$ of space group. The band with red color has mirror eigenvalue
-1, while the bule one has mirror eigenvalue +1. (d) Closeup of band
structure around H point near the Fermi energy as marked in panel (b). The
red color denotes the mirror eigenvalue $-i$ while the green one denotes the
mirror eigenvalue $i$. (e) and (f) are the band structures around H point
with the mirror symmetry broken by shifting Nb atom slightly away from the
equilibrium position. (e) without SOC and (f) with SOC.}
\end{figure*}

\section{III. Results for InNbS$_{2}$}

Here we discuss our band structure results for InNbS$_{2}$. We perform its
density functional GGA\ calculation by using a full potential linear muffin
tin orbital (FP--LMTO) method \cite{FPLMTO} and also cross check the results
with linearized augmented plane wave method as implemented in WIEN2K package 
\cite{WIEN2K}. Both methods provide identical electronic structures. The
orbital character analysis shows that 3$s$ and 3$p$ bands of S atoms are
mainly located at -14 to -12 eV and -7 to -1 eV, respectively. This
indicates that the S--3$s$ and S--3$p$ orbitals are almost completely
filled. Nd--5$d$ states, which are mainly located between -1 and 4 eV, have
also a spectral weight between -7 and -1 eV, indicating a considerable
hybridization between Nb and S. On the other hand, In--5$p$ bands are
distributed mainly above -1 eV. As shown in Fig.2(a), the bands around the
Fermi level are mainly contributed by Nb--5$d_{x^{2}-y^{2}}$/5$d_{xy}$ and
In--6$p_{x}$/6$p_{y}$ states. The Nb--5$d_{x^{2}-y^{2}}$/5$d_{xy}$ bands are
higher in energy than the In--6$p_{x}$/6$p_{y}$ states, however, there is a
band inversion around H point as shown in Fig. 2(c). This band inversion has
also been confirmed by the modified Becke-Johnson (mBJ) exchange potential
calculations \cite{mBJ}. Since L--H--A plane possess mirror symmetry, the
In--6$p_{x}$/6$p_{y}$ (Nb-5$d_{x^{2}-y^{2}}$/5$d_{xy}$) states around H
point can be classified in terms of the mirror eigenvalues $-1$ ($+1$), as
shown in Fig. 2(c). Combining with the time reversal symmetry, this band
inversion guarantees a nodal line in the L--H--A plane \cite{TlTaSe2,PbTaSe2}%
. The schematics of the nodal lines in InNbS$_{2}$ is shown in Fig. 1(c).

\begin{figure*}[tbh]
\center\includegraphics[width=7in, height=5in]{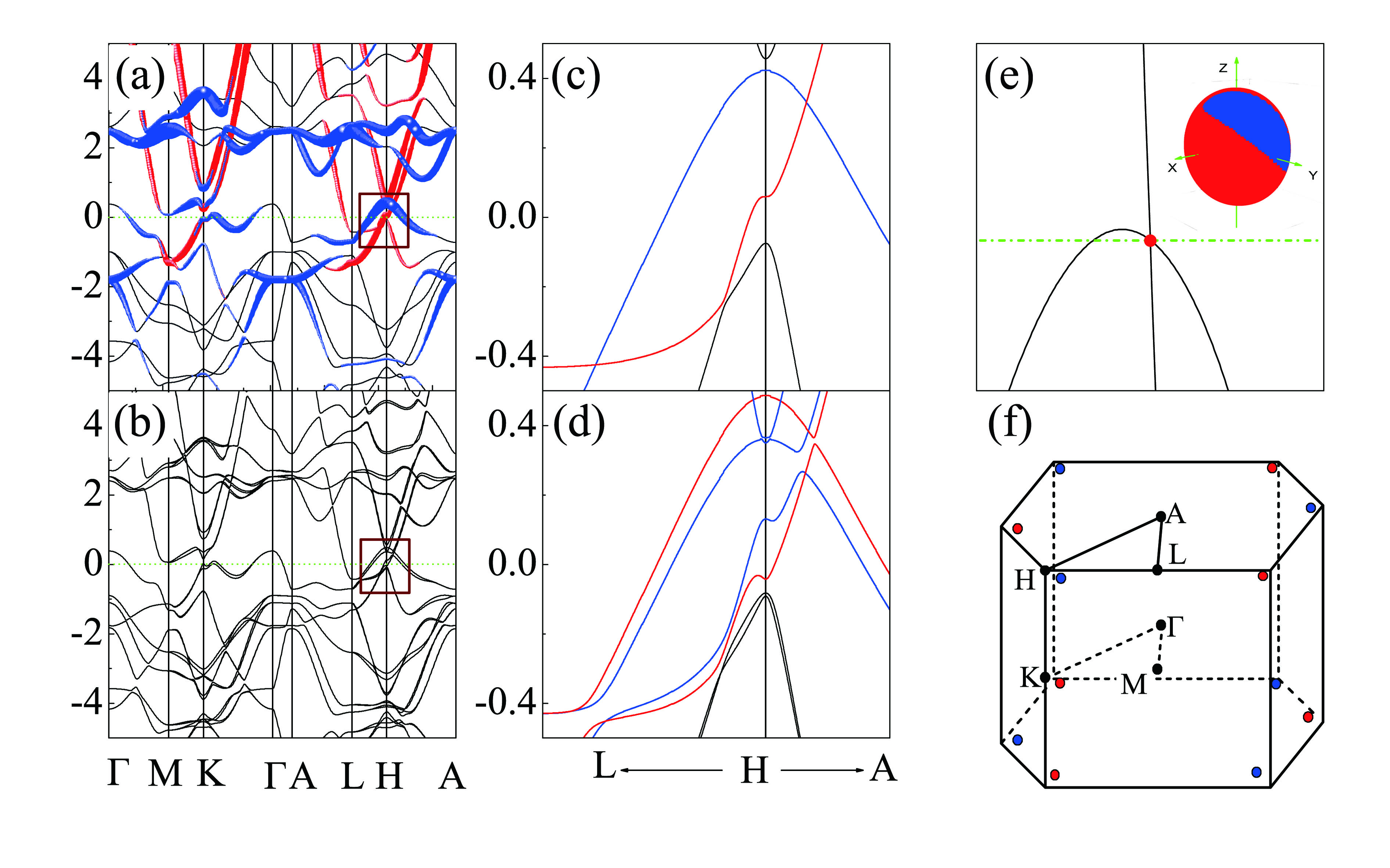}
\caption{(color online). (a) Bulk band structure of InNbSe$_{2}$ without
SOC. The weights of In-6$p_{x}$/6$p_{y}$ (Nb-5d$_{x^{2}-y^{2}}$/5d$_{xy}$)
states are proportional to the width of red (blue) curves. (b) Bulk band
structure of InNbSe$_{2}$ including SOC. (c) Closeup of band structure
around H point as marked in panel (a). The red denote mirror eigenvalue +1,
while the blue denote mirror eigenvalue -1. (d) Closeup of band structure
around H point as marked in panel (b). The red denote mirror eigenvalue +$i$%
, and the blue denote mirror eigenvalue -$i$. (e) band structure around Weyl
point. The insert is the value of $|T(k)|/|U(k)|$, the magnetic field within
the blue area (i.e. $|T(k)|/|U(k)|>1$) can induce negative
magnetoresistence, while the red one ($|T(k)|/|U(k)|<1$) cannot have
negative magnetoresistence\protect\cite{WTe2}. (f) The schematic of the Weyl
points in the first Brillouin Zone. Blue (red) color denote the Chern number 
$+1$($-1$).}
\end{figure*}

To clarify the origin of the band inversion at H point, we calculate the
electronic structure of InNbS$_{2}$ by applying an in--plane tensile strain.
We denote the magnitude of the in--plane strain by $(a-a_{0})/a_{0}$, where $%
a$ and $a_{0}$ denote lattice parameters of the strained and unstrained
systems, respectively. Our calculation reveals that the energy difference
between In--6p$_{x}$/6p$_{y}$ state and Nb--5d$_{x^{2}-y^{2}}$/5d$_{xy}$
states decreases as the in--plane tensile strain increases, and when the
in--plane strain becomes larger than 7\%, the band inversion at H point
disappears. Therefore, the band inversion originates from the crystal field
effect instead of SOC.

\begin{figure*}[tbp]
\includegraphics[width=3.5in, height=3.5in]{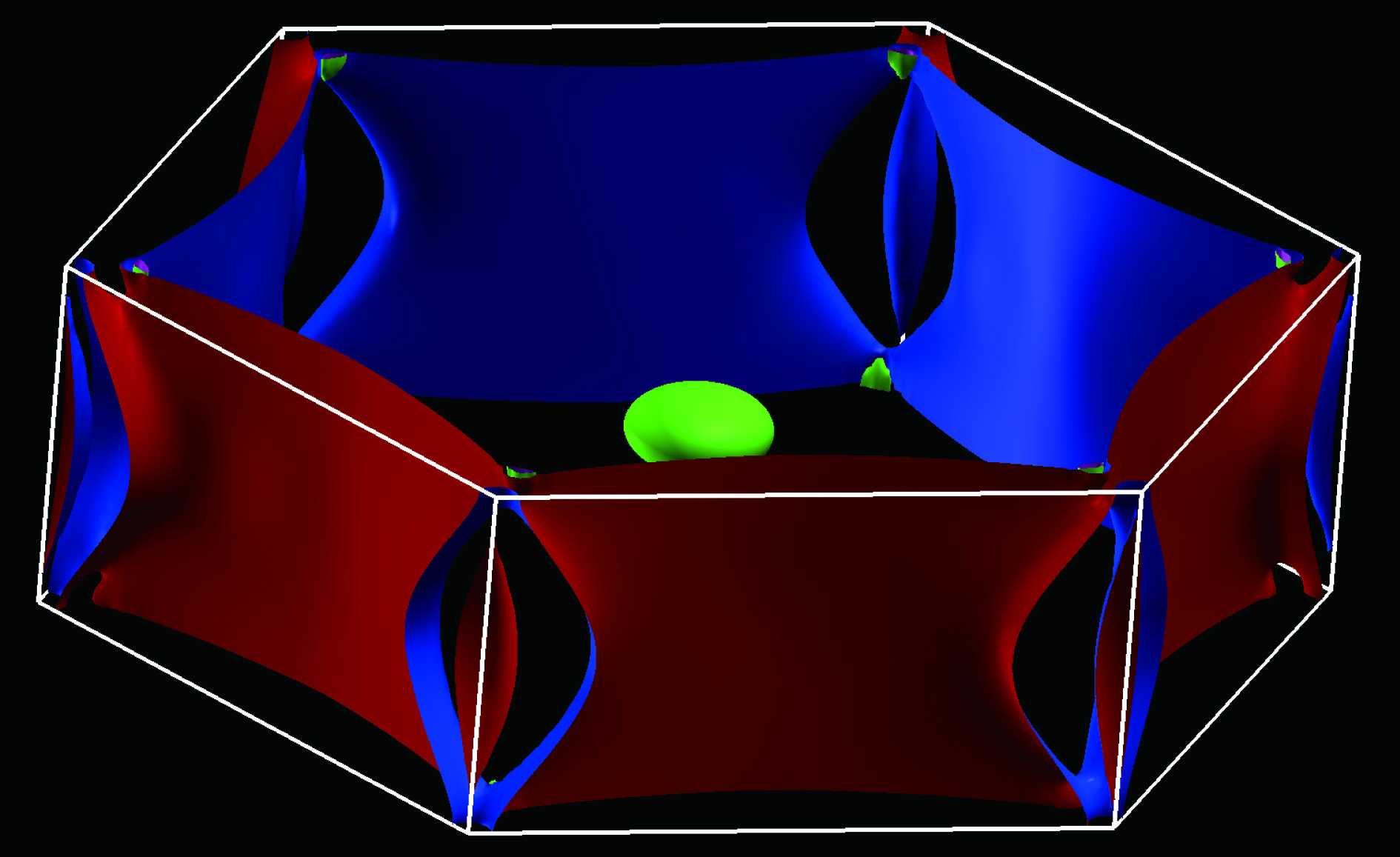} %
\includegraphics[width=3.5in, height=3.5in]{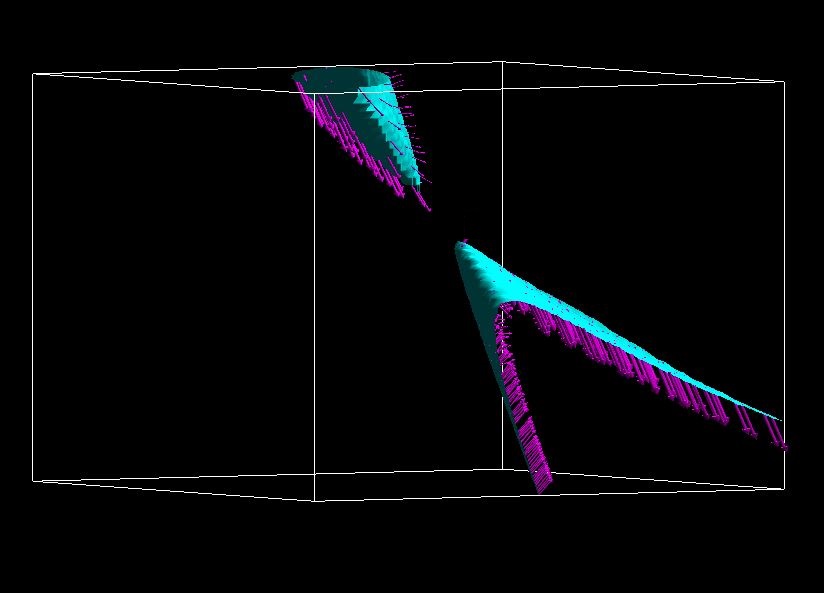}
\caption{(color online). Calculated constant energy surface with a slightly
shifted position (by 0.2 eV up) of the Fermi level for stoichiometric InNbSe$%
_{2}$ showing type II character of the predcited Weyl semimetal: (a) the
green--magenta surface is the hole pocket while the red--blue one is the
electron pocket. (b) The same but in the immidiate vicinity of the Weyl
point. Arrows show spin directions of the one--electron states.}
\end{figure*}

To explore the role of the mirror symmetry, we break it by shifting a Nb
atom by 0.01$\mathring{A}$ along $z$ direction. Without mirror symmetry, the
In--6$p_{x}$/6$p_{y}$ and Nb--5$d_{x^{2}-y^{2}}$/5$d_{xy}$ states belong to
the same irreducible representation and can hybridize with each other. Thus,
the band crossing around the H point becomes gapped as shown in Fig. 2(e).
This clearly demonstrates that the nodal line in InNbS$_{2}$ is indeed
protected by the mirror symmetry.

As a relativistic effect, SOC always exists. Thus we also perform the
calculation to check the effect of SOC. Without inversion, SOC splits each
band into two branches as shown in Fig. 2(b). As a result, there are
spinfull bands near the Fermi level as shown in Fig.2(d). Due to the mirror
symmetry at L--H--A plane, we can classify these four bands by mirror
eigenvalues $\pm i$. Two red bands have mirror eigenvalues $-i$, while blue
ones have mirror eigenvalues $i$. Since SOC\ does not eliminate the band
inversion, there are two separate Weyl rings around the H point. This unique
two separate nodal lines can have unique surface state as discussed in Refs. 
\cite{TlTaSe2,PbTaSe2,Ca3P2}. When mirror symmetry is broken, four bands
around H point are found to belong to the same irreducible representation.
Consequently the Weyl rings are gapped out as shown in Fig. 2(f). This again
shows that the mirror symmetry play a key role in protecting the nodal lines
in InNbS$_{2}$.

\section{IV.\ Results for InNbSe$_{2}$}

We now discuss our band structure calculation for InNbSe$_{2}.$ The results
without and with SOC are shown in Figs. 3(a) and (b), respectively. Our
orbital--character analysis reveals that the states near the Fermi level are
mainly contributed by Nb--5d$_{x^{2}-y^{2}}$/5d$_{xy}$ (denoted by red
color) and In--6p$_{x}$/6p$_{y}$ (denoted by blue color). However, both GGA\
and mBJ\ calculations predict that the energy of Nb--5d$_{x^{2}-y^{2}}$/5d$%
_{xy}$ is higher than In--6p$_{x}$/6p$_{y}$ at the H point. As a result,
there is also a band inversion around the H point, like the case of InNbS$%
_{2}$. This is shown in Fig. 3(a). Again, based on the mirror symmetry, the
bands around the H point can be classified by their mirror eigenvalues. As
shown in Fig. 3(c), the band with a mirror eigenvalue -1 (+1) is marked by
blue (red). Thus, similar with InNbS$_{2}$, time reversal symmetry and the
band inversion result in a nodal line around the H point at the mirror plane.

When SOC is included in the calculation, two crossing bands around the H
point split into four bands as shown in Fig. 3(b). The detailed band
structure around the H point marked in Fig. 3(b) is shown in Fig. 3(d).
Different from the situation in InNbS$_{2}$, the mirror eigenvalues of the
two bands near the Fermi energy are the same (i.e. $-i$). Thus, these bands
are hybridized and open a gap. There are no more nodal lines in this
material. Interestingly, our further calculation shows that InNbSe$_{2}$
becomes a Weyl semimetal with 12 Weyl nodes in the first BZ, as shown in
Fig. 3(e) and (f). These 12 Weyl points are related with each other by
crystal symmetry, consequently they have the same energy. The Weyl points
are searched for by scanning the whole BZ.

To confirm the existence of the Weyl points in InNbSe$_{2}$, we perform
Berry curvature integration based on a computational scheme proposed by
Fukui \textit{et al}\cite{Chern-1}. We define a small cubic region
surrounding each Weyl point. A quantity $\gamma _{P_{l},s}^{n}$, which is
often called the field strength, is defined as\cite{Chern-1,Chern-2}:

\begin{widetext}
\begin{equation*}
\gamma _{P_{l},s}^{n}=\func{Im}\log (\langle n(\mathbf{k}_{l},s)|n(\mathbf{k}%
_{l}+\mathbf{u}_{1},s)\rangle \langle n(\mathbf{k}_{l}+\mathbf{u}_{1},s)|n(%
\mathbf{k}_{l}+\mathbf{u}_{1}+\mathbf{u}_{2},s)\rangle \langle n(\mathbf{k}%
_{l}+\mathbf{u}_{1}+\mathbf{u}_{2},s)|n(\mathbf{k}_{l}+\mathbf{u}%
_{2},s)\rangle \langle n(\mathbf{k}_{l}+\mathbf{u}_{2},s)|n(\mathbf{k}%
_{l},s)\rangle )
\end{equation*}%
\end{widetext}

where $\mathbf{k}_{l}$ is a vector at $l$th mesh point, s=1--6 denotes each
of six faces of the cube, $\mathbf{u}_{1}$ and $\mathbf{u}_{2}$ are vectors
between nearest mesh points for the two directions of the $\mathbf{k}$
vector on the surface of the cube, $P_{l}$ is $l$th smallest closed path
passing by the points $\mathbf{k}_{l}$ and its nearest mesh point. In this
formula, the Chern number is given by the sum over coarse mesh of phases $%
\gamma _{P_{l},s}^{n}$: $C_{n}=\underset{P_{l},s}{\sum }\gamma
_{P_{l},s}^{n} $. The Bloch wave functions $|n(\mathbf{k}_{l},s)\rangle $
are obtained from our first--principles calculations. We employ the 15$%
\times $15 k-mesh on each of the six faces of the cube, which we found to be
sufficient for numerical convergence. We calculate the Chern number of the
Weyl point located at (0.298, 0.298, 0.444) using this method and obtain the
numerical result equal to $+1$. The location of this and other Weyl points
is schematically shown in Fig. 3(f).

We notice that the Weyl points here exist at the boundaries between electron
and hole pockets, therefore the compound can be classified as a type II\
WSM. Our Fig. 3(e) showing a detailed band dispersion in the vicinity of one
Weyl point is very similar to the case of WTe$_{2}$\cite{WTe2} where this
new type of Weyl points has been recently introduced. They appear due to the
tilting term in the linear Weyl Hamiltonian which has led to a finer
classification of topological semimetals\cite{WTe2}. Around the Weyl points,
the energy spectrum can been written as:\textbf{\ }$\varepsilon _{\pm
}(k)=T(k)\pm U(k)$\textbf{.} As a result,\textbf{\ }we expect that InNbSe$%
_{2}$\ will display negative magnetoresistence related to chiral anomaly
only when the direction of the magnetic field falls within the cone where $%
|T(k)|/|U(k)|>1$\ \cite{WTe2} (i.e. the blue area in Fig. 3(e)) \textbf{. }%
We monitor the contact between electron and hole pockets in InNbSe$_{2}$ by
computing the constant energy surface with a slightly shifted position (by
0.2 eV up) of the Fermi level for stoichiometric compound. The result is
shown in Fig. 4(a) where the green--magenta surface is the hole pocket while
the red--blue one is the electron pocket. We can see that the hole pocket
almost touches the electron pocket near the position of the Weyl point.
Fig.4(b) shows the same in the immediate vicinity of the Weyl point together
with the spin distribution of electronic states shown by arrows. For an
ideal Weyl Hamiltonian, the spins are either parallel or antiparallel to the
velocities corresponding to the positive/negative chiralities. In real
compounds such as InNbSe$_{2}$, considered here, this becomes only
approximate, and as seen in Fig. 4(b), spins show a rather high degree of
anisotropy.

\section{V. Fermi Arcs and Effect of Surface Disorder}

In order to examine the Fermi arc surface states of InNbSe$_{2}$ we
determine the one--electron energy bands of 6 unit--cell (24 atomic layers)
slab structure using the full potential linear muffin--tin orbital
(FP--LMTO) method\cite{FPLMTO}. The slab is extended along (001) direction
and terminated by Se atomic layer\ at the top and by In atomic layer at the
bottom. The spacing between the slabs is set to 12\AA . The distance between
In and Se atoms is largest in the original unit cell which together with the
quasi--two--dimensionality of the crystal structure prompts that this should
be most easily cleavable surface in an experimental setup.

\begin{figure*}[tbp]
\includegraphics[width=5in, height=5in]{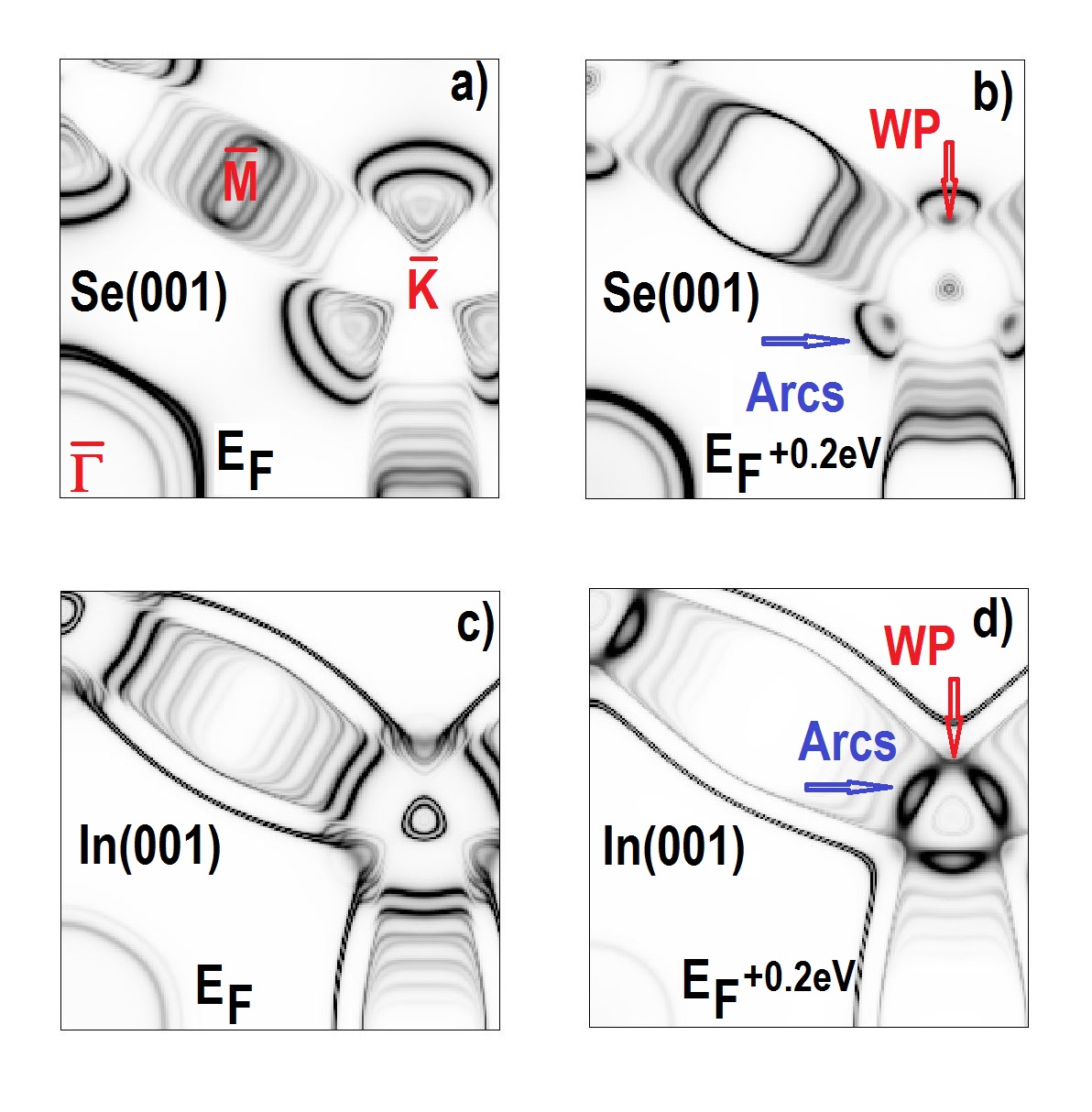}
\caption{(color online). Surface Fermi states of the InNbSe$_{2}$ slab for
the Se terminated (a,b) and In terminated (c,d) (001) surfaces. The position
of the Fermi level in (a) and (c) corresponds to the stoichiometric
compound. Plots (b) and (d) show Fermi surfaces corresponding to the Fermi
level shifted by 0.2 eV when it is tuned to the Weyl points to better
visualize the appearance of the Fermi arcs.}
\end{figure*}

In order to compute the surface Fermi states, we compute surface projected
imaginary Green functions 
\begin{equation*}
\func{Im}G(\mathbf{k},E)=\func{Im}\sum_{j}\frac{\langle \mathbf{k}j|\hat{P}%
_{s}|\mathbf{k}j\rangle }{E-E_{\mathbf{k}j}-i\delta }
\end{equation*}%
where we set $\delta =0.001Ry$, and where the surface projector operator $%
\hat{P}_{s}$ is chosen as a sum over 4 top/bottom atomic layers $\tau $ 
\begin{equation*}
\hat{P}_{s}=\sum_{lm\tau }|\phi _{lmt}\rangle \langle \phi _{lm\tau }|.
\end{equation*}%
Here $\phi _{lm\tau }$ are the solutions of the radial Schroedinger equation
inside a muffin--tin sphere of atom $\tau $ taken with the spherically
symmetric part of the potential\cite{phidot}.

Fig. 5 shows the result of our calculation, where we visualize $\func{Im}G(%
\mathbf{k},E)$ as a function of $\mathbf{k}$ by a color (white is 0 and
black is $1$/$\delta =1000$) within a part of the planar Brillouin Zone
corresponding to the (001) surface unit cell. We distinguish cases for Se
terminated (plots a,b) and In terminated (plots c,d) surfaces. Since the
Weyl points are located not exactly at the Fermi level, we plot $\func{Im}G(%
\mathbf{k},E)$ for the energy $E=E_{F},$ (plots a, c) as well as for the
energy $E=E_{F}+0.2eV$ (plots b, d) which corresponds to the location of the
Weyl points. We note that although (001) surface should be easy to cleave,
the chosen atomic configuration assumes that the Weyl points of opposite
chiralities project onto the same $\mathbf{k}$--point in the surface BZ. It
means that the Fermi arcs extending between opposite chiral charges can
potentially start and end at the same projected Weyl point. We found this to
be the case for the Se terminated surface where small arcs are clearly
visible especially on Fig.5(b) corresponding to the position of the Fermi
level tuned to the Weyl point. The situation is more complicated for the In
terminated surface where there are essentially two lines that are resolved
as connecting the Weyl points on Fig.5(d). We interpret one line to be
potentially the Fermi arc and another one to be either a regular surface
state or a bulk Fermi state projected to the surface BZ.

In a recent work \cite{Disorder} we argued, based on a simulation of a
tight--binding model, that the Fermi arcs should be more surface disorder
tolerant than the regular surface states especially in the vicinity of the
Weyl points where the arcs electronic wave functions are extended well into
the bulk and become less sensitive to the surface disorder. We also found
that the\ particular sensitivity to the surface disorder depends on the
shape of the Fermi arc with the straight arc geometry showing its most
disorder tolerance. Surface disorder is inevitable in a real experimental
setting with vacancies being its primary source. It is therefore interesting
to examine this effect in our proposed InNbSe$_{2}$ WSM.

In order to perform simulation of vacancies on the surface\ of InNbSe$_{2}$,
we use a combination of DFT with a coherent potential approximation (CPA)%
\cite{CPA}, a self--consistent method that allows to extract disorder
induced self--energies $\hat{\Sigma}_{CPA}(E)$ from the FP--LMTO\
calculation. Our recent implementation within FP\ LMTO\ method is described
in Ref. \cite{Disorder}. Figure 6(a,b) shows evolution of the one--electron
Fermi states of the InNbSe$_{2}$ slab structure that are projected onto the
Se terminated surface for two concentrations, $x$=0.05 and $x$=0.1,
respectively, of substitutional vacancies that we impose at its topmost Se
layer. For both concentrations, the Fermi arcs are still visible and
increasing the disorder results in broadening the arcs especially in the
regions always from the Weyl points. Since the arcs electrons are
continuously connected to the bulk Weyl points, the area in the vicinity of
the Weyl points is less affected by disorder. We can contrast this behavior
with the regular surface states which we expect to be more susceptible to
surface disorder. We do not impose any bulk disorder in this calculation,
therefore the bulk states projected onto the surface BZ are unaffected by
the surface vacancies.

\begin{figure*}[tbp]
\includegraphics[width=5in,height=5in]{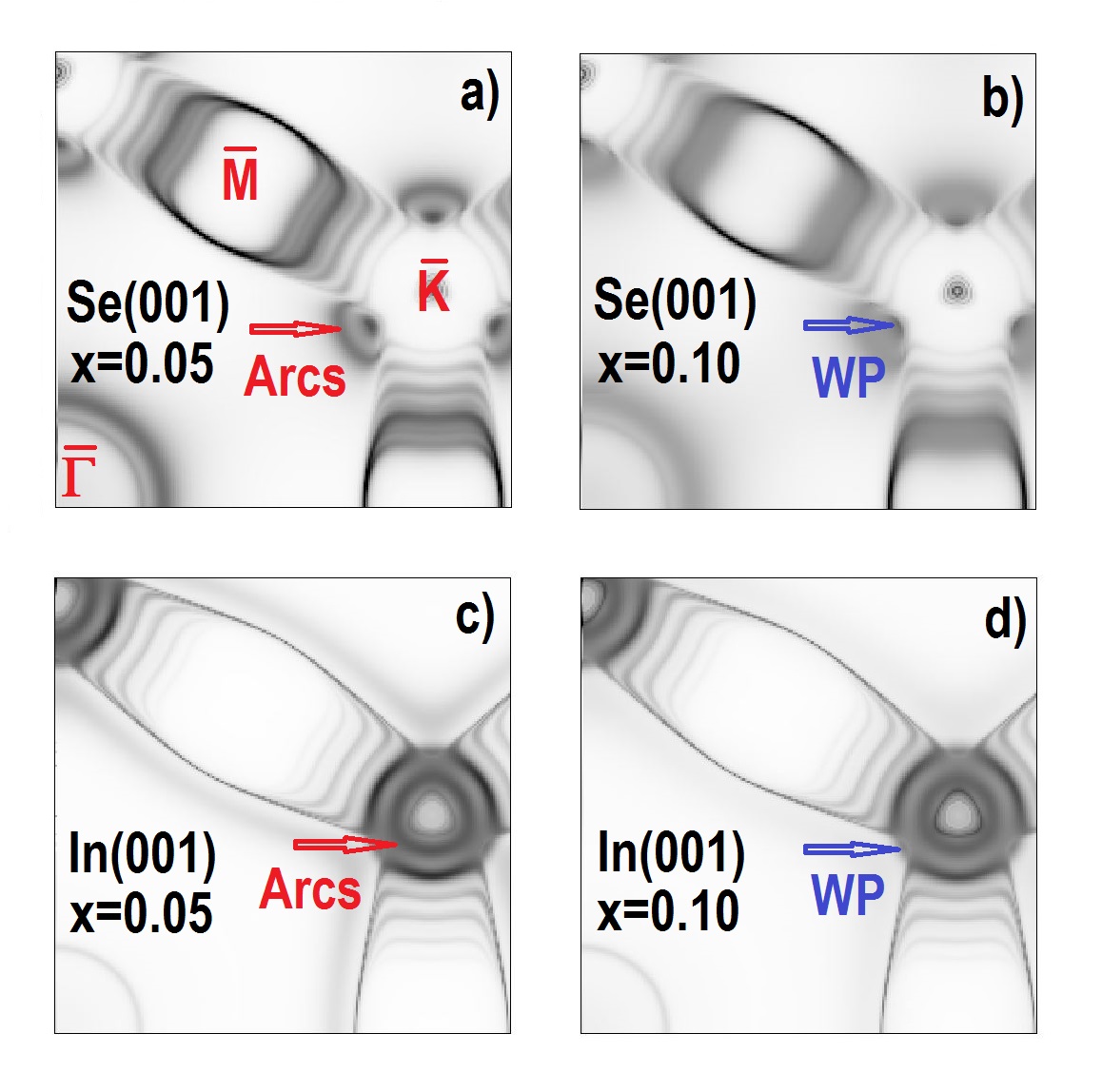}
\caption{(color online). Effect of surface vacancies on the surface Fermi
states of the InNbSe$_{2}$ slab for the Se terminated (a,b) and In
terminated (c,d) (001) surfaces. Plots (a) and (c) correspond to 5\% of
vacancies imposed at the top Se and bottom In layer, respectively. Plots (b)
and (d) correspond to 10\% of the vacancies. The Fermi level is tuned to the
position of the Weyl points to better illustrate the degradation of the
Fermi arc states.}
\end{figure*}

The situation is more complicated for the Fermi arcs appeared at the In
terminated (001) surface. Figure 6(c,d) shows their behavior for $x$=0.05
and $x$=0.1, respectively. We notice that the arcs that connect the Weyl
points in Fig 5(d) slightly change their shape with disorder which possibly
connected to the effect induced by the real part of the disorder
self--energy $\func{Re}\Sigma _{CPA}(E=E_{F}).$ One state is seen in this
simulation to get closer to the $\bar{K}$ point of the surface BZ while
another state has its shape resembling a regular bulk Fermi state projected
onto the surface BZ. Both states acquire much less broadening if we compare
them with other surface states that broaden a lot and almost disappear when
x=0.1. As these arcs show a lot less curvature than the arcs resolved at Se
terminated surface, we therefore speculate that this is likely the effect of
the disorder tolerance for the straight Fermi arcs that we proposed in our
recent work \cite{Disorder}.

\section{VI. Summary}

In summary, by using first--principles calculations, we investigated
topological properties of InNbS$_{2}$ and InNbSe$_{2}$. Our theoretical
analysis showed that InNbS$_{2}$ is a nodal line semimetal even with the
included spin orbit coupling as long as the mirror symmetry preserved. This
significant feature is different from previously proposed materials which
normally neglects SOC. InNbSe$_{2}$ is proposed to be a Type II WSM with 12
Weyl nodes at the same energy level. We also studied the Fermi arcs surface
states and their tolerance to the surface disorder effects. These two
compounds are quasi--2D and easy to cleave, therefore can potentially serve
as an interesting platform for further experimental studies of their
topological electronic states.

\section{Acknowledgement}

The work was supported by National Key R\&D Program of China (No.
2017YFA0303203), the NSFC (No. 11525417, 11374147, 51572085), National Key
Project for Basic Research of China (Grant No. 2014CB921104), the Natural
Science Foundation of Jiangsu Province (Grants No. BK20170821), the Priority
Academic Program Development of Jiangsu Higher Education Institutions. S.S.
was supported by NSF DMR (Grant No. 1411336).

\end{document}